\documentclass[prl,twocolumn,showpacs,superscriptaddress]{revtex4}

\usepackage{amsmath,amssymb}
\usepackage{verbatim}
\usepackage{graphicx}

\usepackage[
      colorlinks=true,
      linkcolor=blue,
      urlcolor=cyan,
      filecolor=magenta,
      citecolor=red,
      pdfstartview=FitV,
      pdftitle={},
        pdfauthor={Arjun Bagchi, Rudranil Basu, Daniel Grumiller, Max Riegler},
        pdfsubject={},
        pdfkeywords={},
        pdfpagemode=None,
        bookmarksopen=true
      ]{hyperref}

\DeclareFontFamily{OT1}{rsfs}{}
\DeclareFontShape{OT1}{rsfs}{m}{n}{ <-7> rsfs5 <7-10> rsfs7 <10->rsfs10}{} 
\DeclareMathAlphabet{\mycal}{OT1}{rsfs}{m}{n}

\newcommand{\e}{\epsilon}

\newcommand{\be}[1]{ \begin{equation}\label{#1} }
\newcommand{\ee}{\end{equation}}
\newcommand{\bea}[1]{\begin{eqnarray}\label{#1} }
\newcommand{\eea}{\end{eqnarray}}
\newcommand{\p}{\partial}

\newcommand{\refb}[1]{(\ref{#1})}

\newcommand{\eq}[2]{\begin{equation} #1 \label{#2} \end{equation}}

\DeclareMathOperator{\extdm}{d}
\newcommand{\extd}{\extdm \!}

\begin{document}

\title{Entanglement entropy in Galilean conformal field theories and flat holography}

\author{Arjun Bagchi}
\email{a.bagchi@iiserpune.ac.in}
\affiliation{Indian Institute of Science Education and Research (IISER), Dr Homi Bhabha Road, Pashan, Pune, Maharashtra 411008, India}
\affiliation{Institute for Theoretical Physics, Vienna University of Technology, Wiedner Hauptstrasse 8--10/136, A-1040 Vienna, Austria}

\author{Rudranil Basu}
\email{rudranil@iiserpune.ac.in}
\affiliation{Indian Institute of Science Education and Research (IISER), Dr Homi Bhabha Road, Pashan, Pune, Maharashtra 411008, India}

\author{Daniel Grumiller}
\email{grumil@hep.itp.tuwien.ac.at}
\affiliation{Institute for Theoretical Physics, Vienna University of Technology, Wiedner Hauptstrasse 8--10/136, A-1040 Vienna, Austria}

\author{Max Riegler}
\email{max.riegler@yukawa.kyoto-u.ac.jp}
\affiliation{Institute for Theoretical Physics, Vienna University of Technology, Wiedner Hauptstrasse 8--10/136, A-1040 Vienna, Austria}
\affiliation{Yukawa Institute for Theoretical Physics (YITP), Kyoto University, Kyoto 606-8502, Japan}

\date{\today}

\preprint{TUW--14--xx}

\begin{abstract} 
We present the analytical calculation of entanglement entropy for a class of two dimensional field theories governed by the symmetries of the Galilean conformal algebra, thus providing a rare example of such an exact computation. These field theories are the putative holographic duals to theories of gravity in three-dimensional asymptotically flat spacetimes. We provide a check of our field theory answers by an analysis of geodesics. We also exploit the Chern--Simons formulation of three-dimensional gravity and adapt recent proposals of calculating entanglement entropy by Wilson lines in this context to find an independent confirmation of our results from holography.
\end{abstract}

\pacs{03.65.Ud, 04.70.Dy, 
 11.15.Yc, 11.25.Hf, 11.25.Tq}

\maketitle

\paragraph{Introduction.}
Entanglement --- originally coined ``Verschr\"ankung'' by Schr\"odinger \cite{Schrodinger:1935} --- is one of the most  fascinating and mysterious features of quantum mechanics. It has been the basis of development of new branches of physics like quantum information and communication \cite{Zeilinger:2000}. For these applications, as well as for theoretical analyses, it is useful to quantify the amount of entanglement of a quantum system, and various measures have been proposed, see e.g.~\cite{Plenio:2007zz}. From a theoretical perspective particularly the entropy of entanglement, or entanglement entropy (EE), has emerged as a valuable tool.

The calculation of EE for interacting quantum field theories (QFTs) remains a daunting task. One of the few examples where analytical calculations can be performed is the EE in two-dimensional (2d) conformal field theories (CFTs) \cite{Calabrese:2004eu}. Here one can invoke the infinite dimensional symmetries of the Virasoro algebra --- the algebra that generates 
conformal transformations in 2d --- to simplify the calculations. E.g.~the EE of a strip of length $\ell$ of a 2d CFT with equal left and right central charges $c$ on the infinite line is given by \cite{Holzhey:1994we,Vidal:2002rm,Calabrese:2004eu}
\eq{
S^{\mbox{\tiny{CFT}}_2}_{\mbox{\tiny{EE}}} = \frac{c}{3}\, \ln{\frac{\ell}{a}} 
}{eq:EE}
where $a$ is the lattice size that regulates the ultraviolet divergences. The problem of calculating EE in QFTs quickly becomes intractable with increasing dimension and complexity, since in general there is no infinite dimensional symmetry algebra to fall back on.

The holographic principle has played a vital role in the recent resurgence of EE calculations for QFTs. According to this principle, a theory of (quantum) gravity in a certain spacetime is completely equivalent to a QFT without gravity living on the boundary of that space-time. The celebrated Anti-de~Sitter/conformal field theory (AdS/CFT) correspondence has given this bold proposal a firm footing in the context of string theory by relating Type IIB superstring theory on AdS$_5 \otimes$S$^5$ to $\mathcal{N} =4$ supersymmetric Yang-Mills theory on its boundary \cite{Maldacena:1997re}. 

As a remarkable consequence of holography, the calculation of EE in CFTs has been conjectured to be equivalent to the computation of the area of an extremal co-dimension-two-surface in AdS \cite{Ryu:2006bv}. There are recent developments which have built towards a proof of this relation \cite{Fursaev:2006ih,Klebanov:2007ws,Lewkowycz:2013nqa}. Thus, using holography, the involved calculation of EE in higher dimensional CFTs has boiled down to the fairly simple computation of areas of extremal surfaces in AdS. See \cite{Bombelli:1986rw,Srednicki:1993im,Frolov:1993ym,Kitaev:2005dm,Ryu:2006ef,Solodukhin:2006xv,Hubeny:2007xt,Wolf:2007,Nishioka:2009un,Headrick:2010zt,Casini:2011kv,Almheiri:2012rt,Maldacena:2013xja,Marolf:2013dba,Papadodimas:2013wnh,Harlow:2014yka} for selected pre-cursors and elaborations on the connection between 
gravity and entanglement.

The present work has two main goals: to calculate EE for a novel class of QFTs, and to derive EE for the same class of QFTs holographically. 

Related to the first goal, we shall calculate EE for Galilean conformal field theories (GCFTs), thus providing a rare example of an exact computation of EE.

Concerning the second goal, 2d GCFTs \cite{Bagchi:2009my} were proposed as duals \cite{Bagchi:2010zz} to three-dimensional (3d) gravity in asymptotically flat spacetimes \cite{Barnich:2006av}. There has been considerable interest in trying to understand holography in flat spacetimes in the light of these findings \cite{Bagchi:2012yk, Bagchi:2012xr, Barnich:2012xq, Barnich:2012aw, Bagchi:2012cy, Bagchi:2013bga, Bagchi:2013lma, Afshar:2013vka, Gonzalez:2013oaa, Costa:2013vza, Fareghbal:2013ifa, Krishnan:2013tza, Krishnan:2013wta, Bagchi:2013qva, Detournay:2014fva, Barnich:2014kra, Barnich:2014cwa, Riegler:2014bia, Fareghbal:2014qga}. We shall provide further evidence for a flat space/GCFT correspondence by holographically calculating EE in 3d asymptotically flat spacetimes, thereby confirming the 2d GCFT calculations.

\paragraph{EE in 2d CFTs.} 
Let us now briefly recall the strategy employed for the calculation of EE in 2d CFTs, starting with the definition of EE. If $\rho$ is the density matrix of a quantum system with multiple degrees of freedom, and the Hilbert space of the system can be written as the direct product $\mathcal{H} = \mathcal{H}_A \otimes \mathcal{H}_B$, the reduced density matrix $\rho_A$ for the subsystem $A$ is defined by the partial trace over the Hilbert subspace $\mathcal{H}_B$, $\rho_A = \mbox{Tr}_B\, \rho$. The EE is then the corresponding von Neumann entropy
\eq{
S_A = - \mbox{Tr} \big(\rho_A \ln \rho_A\big)\,.
}{eq:angelinajolie}

EE is conveniently calculated by employing the replica trick, using Renyi entropies, $S_A^{(n)} = \frac{1}{1-n} \ln \mbox{Tr} \rho_A^n$. EE \eqref{eq:angelinajolie} emerges as the limit $S_A=\lim_{n\to 1} S_A^{(n)}$.
More specifically, in a 2d QFT defined on a lattice, where the subsystem $A$ consists of $n$ disjoint intervals, one computes Renyi entropy $S_A^{(n)}$ by making $n$ copies of the system with one disjoint interval and sewing them together. This defines an $n$-sheeted Riemann surface with a partition function $Z_n(A)$ and $\mbox{Tr} \rho^n_A = \frac{Z_n(A)}{Z_1(A)^n}$. From this one can calculate the EE of the system as a limit $S_A = - \lim_{n \to 1} \frac{\p}{\p n} \frac{Z_n(A)}{Z_1(A)^n}$.
For details of this construction see e.g.~\cite{Calabrese:2009qy,Casini:2009sr}. 

We focus here on the EE of a single interval $[u, v]$ of length $\ell = |u-v|$ in an infinitely long one-dimensional quantum system at zero temperature and recapitulate the main insights and results.
One key ingredient is to map the $n$-sheeted Riemann surface $\mathcal{R}_n$ to the complex plane $\mathbb{C}$, where $n$ labels the $n^{\textrm{th}}$ Renyi entropy. Another relevant aspect is that in a 2d CFT, the 
energy-momentum (EM) tensor on $\mathcal{R}_n$ is related to that on $\mathbb{C}$ by $T(w) = \left( \frac{dz}{dw}\right)^2 T(z) + \frac{c}{12} \{z, w \}$ where $c$ is the central charge of the CFT and $\{, \}$ the Schwarzian derivative. Taking expectation values of both sides and exploiting $\langle T(z) \rangle = 0$ on $\mathbb{C}$ yields
\be{Rn}
\langle T(w) \rangle_{\mathcal{R}_n} = \frac{c}{24} \,\Big(1 - \frac{1}{n^2}\Big) \frac{(v-u)^2}{(w-u)^2 (w - v)^2}\,.
\ee
Conformal Ward identities then imply
\be{tward}
\langle T(w) \rangle_{\mathcal{R}_n} = \frac{\langle T(w) \tilde{\Phi}_n (u) \tilde{\Phi}_{-n} (v) \rangle_{\mathbb{C}}}{\langle\tilde{\Phi}_n (u) \tilde{\Phi}_{-n} (v) \rangle_{\mathbb{C}}}
\ee
where $\tilde{\Phi}_n, \tilde{\Phi}_{-n}$ are twist fields in the 2d CFT, viz., primaries with conformal weights $\Delta = \bar{\Delta} =  (c/24)(1 - 1/n^2)$. The key observation is that $\mbox{Tr} \rho^n_A$ transforms under a general conformal transformation as a 2-point function of such primaries. From this, Calabrese and Cardy deduced 
$\mbox{Tr} \rho^n_A \propto \big(\frac{v-u}{a}\big)^{-c (n-1/n)/6}$.
The ensuing expression for Renyi entropy 
\be{eq:sl1}
S_A^{(n)} = \frac{c}{6} \Big( 1+ \frac{1}{n} \Big) \ln \frac{\ell}{a} 
\ee
in the limit $n\to 1$ then establishes the result \eqref{eq:EE} for EE.

\paragraph{GCFT symmetries and correlators.} 
GCFTs are the non-relativistic analogues of CFTs. 
They can also be defined independently from any relativistic parent CFT \cite{Bagchi:2009my}.
We now build towards our goal of calculating EE in GCFTs. A first step is to analyze the symmetry algebra, its highest weight representations and restrictions from Ward identities on correlation functions.
The symmetry generators $L_n$, $M_n$ 
of a 2d GCFT obey 
\bea{eq:sl2}
[L_n,\, L_m] &=& (n-m) L_{n+m} + \frac{c_L}{12}\, (n^3-n)\, \delta_{n+m,\,0} \crcr
[L_n,\, M_m] &=& (n-m) M_{n+m} + \frac{c_M}{12}\, (n^3-n)\, \delta_{n+m,\,0} \crcr
[M_n,\, M_m] &=& 0 \qquad\quad n,m\in\mathbb{Z} \qquad c_L,c_M\in\mathbb{R}\,.
 \label{eq:algebra}
\eea
We consider now the representation theory of this Galilean conformal algebra (GCA) to construct the Hilbert space of a GCFT. We choose to work with the highest weight representation \cite{Bagchi:2009ca, Bagchi:2009pe}. Any state in this field theory is labeled by two weights
$L_0 | h_L, h_M \rangle = h_L | h_L, h_M \rangle$, $M_0 | h_L, h_M \rangle = h_M | h_L, h_M \rangle$.
Following CFT literature, we define primary states by demanding that they be annihilated by all generators $L_n$, $M_n$ with positive $n$.
We shall use the plane representation of the GCA \eqref{eq:algebra}
\be{eq:sl5}
L_n = x^{n+1}\, \p_x + (n+1) x^n y\, \p_y\,, \qquad
M_n = x^{n+1}\, \p_y
\ee
where $x$, $y$ are our 2d coordinates.
We define the EM tensor of a GCFT on a plane \cite{Bagchi:2010vw} by
\begin{align}
T_{(1)} &= \sum_{n} \Big(L_n  + (n+2)\, \frac{y}{x}\, M_n \Big)\, x^{-n-2} \\
T_{(2)} &= \sum_{n} M_n x^{-n-2}\,.
\end{align}
An important result is that the 2-point correlator of primary fields $\{\Phi^i (x_i, y_i)\}$ with weights $h^i_L , h^i_M$ has a particularly nice form in the plane representation, fixed by the symmetries of the GCA (up to normalization):
\be{2pt}
\langle \Phi^1 (x_1, y_1) \Phi^2 (x_2, y_2)\rangle \propto x_{12}^{-2 h_L}\, e^{-2 h_M \frac{y_{12}}{x_{12}}}\, \delta_{h^1_L}^{h^2_L}\, \delta_{h^1_M}^{h^2_M}
\ee
Here and below expressions like $x_{12}$ always denote differences between coordinates, $x_{12}=x_1-x_2$.

The 3-point correlator of the EM tensor with two primary fields (of weight $\{h_L, h_M\}$) is determined from Ward identities and the 2-point correlators \refb{2pt}:
\bea{tpp}
\langle T_{(1)}(x_w,y_w) \Phi(x_1, y_1) \Phi(x_2, y_2)\rangle &=& \Big(\frac{x_{12}}{x_{w1} x_{w2}}\Big)^2 \\
&& \hspace{-5cm} \Big[ h_L - 2 h_M \Big(\frac{y_{12}}{x_{12}} - \frac{y_{w1}}{x_{w1}} - \frac{y_{w2}}{x_{w2}} \Big) \Big]\,  x_{12}^{-2h_L} \, e^ {-2h_M \frac{y_{12}}{x_{12}}} \nonumber\\
\langle T_{(2)}(x_w,y_w) \Phi(x_1, y_1) \Phi(x_2, y_2)\rangle &=& \Big(\frac{x_{12}}{x_{w1} x_{w2}}\Big)^2   \\ && \hspace{-0.5cm} h_M \, x_{12}^{-2h_L}\, e^ {-2h_M \frac{y_{12}}{x_{12}}} \nonumber
\eea
The 3-point correlators above will play a crucial role in our derivation of EE for GCFTs.

\paragraph{EE in 2d GCFTs.} We are now ready to construct the EE for a GCFT of an infinite one-dimensional system for one interval at vanishing temperature.

We intend to use the analogue of \refb{tward} in 2d GCFTs in order to be able to make the statement that also in GCFTs $\mbox{Tr} \rho^n_A$ depends only on the form of the 2-point function of primary operators. We have already computed the right hand side of \refb{tward} from symmetry considerations above. For arriving at the left hand side, we shall employ a limiting technique from 2d CFTs~\footnote{%
At this point we are restricting ourselves to GCFTs that arise as (non-relativistic) limits from relativistic 2d CFTs. It should be possible to understand our analysis from a purely GCFT point of view without recourse to a limit, but at the moment the generalization of the transformation of the EM tensor of the GCFT presents a challenging obstacle, which we choose to by-pass for the purposes of the present problem.
}. We need the GCFT equivalent of \refb{Rn}, assuming  a smooth non-relativistic limit. The expressions for the GCFT EM tensor in terms of the 
original CFT EM tensor read: $T_{(1)}(x_w, y_w)=\lim_{\epsilon \rightarrow 0} \left[T(w)+\bar{T}(\bar{w})\right], \,T_{(2)}(x_w, y_w)=\lim_{\epsilon \rightarrow 0} \epsilon \left[T(w)-\bar{T}(\bar{w})\right]$ with $w= x_w + \epsilon y_w$ and the limit is defined by $\e \to 0$. Denoting $u=x_1+ \epsilon y_1$ and $v=x_2+ \epsilon y_2$, this leads to:
\bea{Texp}
\langle T_{(1)}(x_w, y_w)\rangle _{\mathcal{R}_n} &=& \Big(1-\frac{1}{n^2}\Big) \Big(\frac{x_{12}}{x_{w1} x_{w2}}\Big)^2 \\ &&\Big[ \frac{c_L}{24} - \frac{c_M}{12} \Big(\frac{y_{12}}{x_{12}}-\frac{y_{w1}}{x_{w1}}-\frac{y_{w2}}{x_{w2}}\Big)\Big] \nonumber \\
\langle T_{(2)}(x_w, y_w)\rangle _{\mathcal{R}_n}&=& \Big(1-\frac{1}{n^2}\Big)\Big(\frac{x_{12}}{x_{w1} x_{w2}}\Big)^2 \, \frac{c_M}{24}
\label{eq:lalapetz}
\eea

Combining the results \refb{2pt}-\refb{eq:lalapetz} obtains ($i = 1,2$)
\be{tward1}
\langle T_{(i)}(x, y) \rangle_{\mathcal{R}_n} = \frac{\langle T_{(i)}(x,y) \Phi_n (x_1, y_1) \Phi_{-n} (x_2, y_2) \rangle_{\mathbb{C}}}{\langle\Phi_n (x_1, y_1) \Phi_{-n} (x_2, y_2) \rangle_{\mathbb{C}}}
\ee
and determines the weights of the 2d GCFT twist fields $\Phi_n (x, y)$ as $h_L = \frac{c_L}{24} \left( 1- \frac{1}{n^2}\right)$ and  $h_M = \frac{c_M}{24} \left( 1- \frac{1}{n^2}\right)$. 
Arguments analogous to the ones for 2d CFTs then permit us to infer
\bea{eq:sl6}
\mbox{Tr} \, \rho_A^n = k_n \langle \Phi _{h_L , h_M}(x_1, y_1) \Phi _{h_L , h_M}(x_2, y_2)\rangle ^n _{\mathbb{C}} \nonumber\\
= k_n \, x_{12}^{-\frac{c_L}{12}(n-\frac{1}{n})} \exp \Big[\frac{c_M}{12} (n-\frac{1}{n}) \frac{y_{12}}{x_{12}} \Big]
\eea
for some constants $k_n$. We fix $k_1$ conveniently. 

We derive EE as before from a limit of Renyi entropies, $S_{A} = -\lim_{ n \rightarrow 1} \frac{\partial}{\partial n} \mbox{Tr} \, \rho_A^n $, and finally obtain
\eq{
\boxed{\phantom{\Bigg(}
S^{\mbox{\tiny{GCFT}}_2}_{\mbox{\tiny{EE}}}  = \frac{c_L}{6}\, \ln \frac{\ell_x}{a} + \frac{c_M}{6} \,\frac{\ell_y}{\ell_x}
\phantom{\Bigg(}}
}{eq:GCAEE}
where $\ell_x=a x_{12}$, $\ell_y=a y_{12}$, and $a$ is again the lattice size that regularizes ultra-violet divergences. The EE formula \refb{eq:GCAEE} is one of our main results and completes the first goal, namely to calculate EE for a 2d GCFT. Note the similarity of the first term in the GCFT EE with the expression for the CFT EE \eqref{eq:EE}. This is expected, since in theories with vanishing $c_M$ the non-trivial part of the GCA \eqref{eq:algebra} reduces to a single copy of the Virasoro algebra \cite{Bagchi:2009pe,Bagchi:2012yk}.

\paragraph{Generalization to finite temperature or finite size.}
We generalize now our main result \eqref{eq:GCAEE}, starting with the consideration of finite temperature $T=\beta^{-1}$. We follow again the lead of Calabrese and Cardy \cite{Calabrese:2004eu}.
For this purpose, we utilize the fact that $\mbox{Tr} \, \rho_A^n$ behaves like the 2-point function of some primary fields in a GCFT. We use the transformation laws of the fields from the plane to the cylinder elucidated in \cite{Bagchi:2013qva} to find after some algebra: 
\be{eef}
S =\frac{c_L}{6} \ln \Big(\frac{\beta}{\pi}\sinh \big( \frac{\pi}{\beta} \xi_{12}\big) \Big) +  \frac{\pi c_M}{6 \beta} \rho_{12} \coth \big( \frac{\pi}{\beta} \xi_{12}\big)
\ee
This is the EE for a mixed state in the 2d GCFT at finite temperature $\beta^{-1}$. 

The result \refb{eef} has the expected behaviour under two extreme limits. The first limit, $\xi_{12},\, \rho_{12} \ll \beta$, gives back the answer on the plane \refb{eq:GCAEE} with $\xi_{12}\to \ell_x/a$ and $\rho_{12} \to \ell_y/a$. The second limit, $\xi_{12},\, \rho_{12} \gg \beta$, yields
\eq{
S = \frac{\pi}{6 \beta} \big(c_L \,\xi_{12} + c_M \,\rho_{12} \big) + \frac{c_L}{6}\,\ln\beta + \dots 
}{eq:EElimit2}
where the ellipsis denotes temperature-independent and subleading terms in $\beta$.
To leading order the EE \eqref{eq:EElimit2} is extensive, i.e., linear in the separations, which is completely analog to what happens in 2d CFTs \cite{Calabrese:2004eu}. The subleading, logarithmic, term is universal, depending only on $c_L$.

To find the entropy of a subsystem of length $(\xi_{12}, \rho_{12})$ in a finite system of length $L$ in its ground state, we use a similar analysis, replacing $\beta \to L$ (and orienting the cuts along which the twist operators lie perpendicular to the previous case, which results in some changes). We then obtain a result for EE in systems of length $L$:
\eq{
S=\frac{c_L}{6} \ln \Big(\frac{L}{\pi}\sin \big( \frac{\pi}{L} \xi_{12}\big) \Big) +  \frac{\pi c_M}{6 L} \rho_{12} \cot \big( \frac{\pi}{L} \xi_{12}\big) 
}{eq:FSEE}

\paragraph{Towards higher dimensions.}
One of the great advantages of GCFTs is that the infinite dimensional symmetry extends to all spacetime dimensions. These have been realized in physical systems e.g.~hydrodynamics \cite{Bagchi:2009my} and more recently in Galilean electromagnetic theories \cite{Bagchi:2014ysa}. It is plausible that the constructions above would work similarly for higher dimensional cases. 

\paragraph{Holographic EE in GCFT.}
In the remainder of our work we focus on the second goal, namely to derive our results for EE holographically. As a first step, we summarize aspects of proposed 3d gravity duals to 2d GCFTs. A minimal requirement for a putative gravity dual is that its asymptotic symmetries match the symmetries of the corresponding QFT. Flat space holography in 3d realizes this requirement \cite{Barnich:2006av,Bagchi:2010zz}. (See \cite{Li:2010dr} for a different approach to holography and entanglement in flat space.)

\paragraph{Flat space geodesics.}
Following the Ryu--Takayanagi prescription \cite{Ryu:2006bv} one can determine EE holographically in 3d (Einstein) gravity by  computing the length of geodesics on equal time slices. 
To determine the relevant geodesics we consider the most general zero-mode solutions of 3d flat space Einstein gravity, given by the line-element \cite{Barnich:2012aw}
\eq{
\extd s^2 = M\,\extd u^2 - 2\extd u\extd r + J\extd u\extd\varphi + r^2\extd\varphi^2
}{eq:FSC}
with mass parameter $M$ and angular momentum parameter $J$. For $M=J=0$ the solution is known as null orbifold \cite{Horowitz:1990ap,FigueroaO'Farrill:2001nx,Liu:2002ft,Simon:2002ma}. For $M=-1$ and $J=0$ the solution is (global) flat space. The more generic case $M\geq 0$ and $J\neq 0$ corresponds to flat space cosmology solutions \cite{Cornalba:2002fi,Cornalba:2003kd}, which we will not consider in the current work.

The central charges for flat space Einstein gravity are $c_L=0$ and $c_M=3/G_N$, where $G_N$ is the 3d Newton constant \cite{Barnich:2006av}. Therefore, the only term that remains in EE for flat space Einstein gravity is proportional to $\ell_y$ [$\rho_{12}$] on the plane at zero temperature [at finite temperature or for finite size], see \eqref{eq:GCAEE} [\eqref{eef} or \eqref{eq:FSEE}]. In the coordinates above this separation translates into the separation $u_{12}$. This means that taking a constant $u$-slice we should obtain vanishing EE. Indeed, this is precisely what one obtains from calculating the length of such geodesics.

While the conclusion above shows consistency of the holographic calculation with the GCFT results, it is more interesting to consider the general case where both central charges are non-zero. For this purpose, we no longer can consider geodesics of Einstein gravity solutions. 

\paragraph{Holographic EE from Chern--Simons formulation.} 
Einstein gravity in 3d can conveniently be reformulated as a Chern--Simons gauge theory \cite{Achucarro:1986vz,Witten:1988hc} with the action
	\begin{equation}\label{eq:ICS}
		I_{CS}[\mathcal{A}]=\frac{k}{4\pi}\int\langle\mathcal{A}\wedge\extd\mathcal{A}+\tfrac{2}{3}\mathcal{A}\wedge\mathcal{A}\wedge\mathcal{A}\rangle,
	\end{equation}
where $k$ is the Chern-Simons level, $\mathcal{A}$ takes values in some gauge algebra $\mathfrak{g}$ and $\langle\ldots\rangle$ denotes a suitable invariant bilinear form on the gauge algebra. In flat space the gauge algebra is given by $\mathfrak{g}=\mathfrak{isl}(2)$, which is precisely the algebra \eqref{eq:algebra} restricted to $n=0,\pm 1$. 
The flat space connection describing spacetimes \eqref{eq:FSC} reads \cite{Gary:prep}
\begin{gather}
\label{eq:sl42}
\mathcal{A}=b^{-1}\big(\extd  + a\big) b\qquad \textrm{with} \qquad b=e^{\frac{r}{2} M_{-1}}\,, \\
a=\Big( M_{1}-\frac{M}{4}M_{-1}\Big)\extd u + \Big(L_{1}-\frac{M}{4}L_{-1} -\frac{J}{4}M_{-1} \Big)\extd\varphi \,. \nonumber
\end{gather}
Concurrent with the Grassmanian interpretation of \cite{Krishnan:2013wta} we refer to the generators $L_n$ ($M_n$) as ``even''  (``odd'').

In \cite{deBoer:2013vca,Ammon:2013hba,Castro:2014tta} it has been argued for AdS$_3$ that using the Chern--Simons formalism holographic EE of a given interval at the boundary of AdS is given by (the logarithm of) a Wilson line attached at the endpoints of this interval.
\eq{
\begin{gathered}
S_{\textrm{\tiny EE}} =-\ln\left[W_\mathcal{R}(C)\right]\\
W_{\mathcal{R}}(C) =\textnormal{Tr}_\mathcal{R}\left(\mathcal{P}e^{\int_C\mathcal{A}}\right)=\int\mathcal{D}Ue^{-S(U;\mathcal{A})_C}
\end{gathered}
}{eq:nolabel}
Here $\mathcal{R}$ denotes a suitable representation of the symmetry algebra at hand, $\mathcal{P}$ stands for path-ordering, $C$ denotes the contour of the Wilson line, and $U$ refers to a topological probe exploring the bulk geometry.

For flat space we assume a split of the probe $U$ and the action $S(U;\mathcal{A})_C$ into even and odd parts labeled by $L$ and $M$, respectively, i.e., $U=U^LU^M$, and $S(U;\mathcal{A})_C=S_L(U_L;A_L)_C+S_M(U_M;A_M)_C$, where $\mathcal{A}=A_L+A_M$.
Using this assumption the even action $S_L$ is given by
\eq{
S_L=\int_C\extd s \Big(\langle P_L(D_LU_L)U_L^{-1}\rangle_L+\lambda_L\big(\langle P_L^2\rangle_L-C_2^L\big)\Big) 
}{eq:SL}
where $D_L=\partial_s+A_L$, $P_L$ is the canonical momentum conjugate to $U_L$, and $\lambda_L$ is a Lagrange multiplier that sets the quadratic Casimir to the right value $C_2^L = c_L^2/288$. Note that at this stage we do no longer assume that we are necessarily dealing with Einstein gravity, where $c_L$ vanishes; an example \cite{Bagchi:2012yk} for a theory with $c_L\neq 0$  is topologically massive gravity \cite{Deser:1982vy} whose AdS EE was derived holographically in \cite{Castro:2014tta}.
The brackets $\langle\ldots\rangle_L$ denote a suitable invariant bilinear form satisfying
$\langle P_L,\,P_L\rangle_L = -P_L^{L_{-1}}P_L^{L_1} + \tfrac{1}{2}P_L^{L_0}P_L^{L_0} - P_L^{L_1}P_L^{L_{-1}}$.
The odd action $S_M$ (and all quantities therein) are obtained from the even action \eqref{eq:SL}, replacing everywhere indices $L\to M$.

We determine now the equations of motions from these two actions and find solutions to them, using the so-called ``nothingness'' trick: we first solve the equations of motion for the trivial case $A_L=A_M=0$ and determine $U_{L,M}^{(0)}$ and $P_{L,M}^{(0)}$. Then we use a (large) gauge transformation
		$U(s) = g(s)U^{(0)}_{L}U^{(0)}_{M}g^{-1}(s)$,
where the group element $g(s)$ is in the component continuously connected with the identity to obtain solutions for non-trivial gauge fields $A_{L,M}$~\footnote{%
Locally, any Chern--Simons connection that solves the field equations is pure gauge and thus trivial. However, globally the connection can be non-trivial and contains information about the physical state, like its mass or angular momentum. By large gauge transformations trivial configurations can be mapped to non-trivial ones.
}. The actions $S_L$ and $S_M$ then reduce on-shell to
$S_{L,M}^{\textnormal{on-shell}}=-2\Delta\alpha_{L,M}C_2^{L,M}$
with $\Delta\alpha_{L,M}=\alpha_{L,M}(s_f)-\alpha_{L,M}(0)$ determined by $U_{L,M}^{(0)}$.

The only thing left to do is to choose suitable boundary conditions for the topological probe $U(s)$ at the  beginning $s=0$ and the end $s=s_f$ of the boundary interval and to determine $\Delta\alpha_{L,M}$. The reason why it is sufficient to determine $\Delta\alpha_{L,M}$ is that  in the semiclassical limit $\int\mathcal{D}Ue^{-S(U;\mathcal{A})_C}\sim e^{-S(U;\mathcal{A})_C}$ and thus
\eq{
S_{{\mbox{\tiny{EE}}}} = -2\Delta\alpha_LC_2^L - 2\Delta\alpha_M C_2^M\,.
}{eq:hEE}
We have found suitable boundary conditions, whose detailed discussion will be presented elsewhere \cite{Bagchi:prep}. 

From the non-trivial connection \eqref{eq:sl42} and its split into even and odd parts, $A_L+A_M=g\extd g^{-1}$ with $g=b^{-1}e^{-\int a_i\extd x^i}$, one can now readily obtain the EE for various interesting spacetimes. We collect these results below. We define $\ell_u= u_f-u_0$ and $\ell_\varphi=\varphi_f-\varphi_0$; the quantity $a$ is again a regularizing cut-off.
Applying the procedure above yields the EE of the null orbifold ($M=J=0$), to be compared with \eqref{eq:GCAEE}.
\eq{
\boxed{
S_{\mbox{\tiny{EE}}}^{\textrm{\tiny null}} = \frac{c_L}{6}\,\ln\frac{\ell_\varphi}{a} + \frac{c_M}{6}\,\frac{\ell_u}{\ell_\varphi}
}
}{eq:EE0}
Similarly, for global flat space ($M=-1$, $J=0$) we obtain the EE [to be compared with \eqref{eq:FSEE} for $L=2\pi$]
\eq{
S_{{\mbox{\tiny{EE}}}}^{\textrm{\tiny flat}} = \frac{c_L}{6}\ln\big(2 \sin\frac{\ell_\varphi}{2}\big) + \frac{c_M}{12}\ell_u\cot\frac{\ell_\varphi}{2}\,.
}{eq:EEflat}
The holographic EE results above are in precise agreement with the ones obtained on the field theory side. 


\acknowledgments

We thank Shamik Banerjee, Alejandra Castro, St\'ephane Detournay, Reza Fareghbal, Michael Gary, Nabil Iqbal, Juan Jottar, Esperanza Lopez, Jan Rosseel, Joan Sim\'on, Stefan Stricker, Tadashi Takayanagi and Erik Tonni for discussions.

AB was supported by an INSPIRE award of the Department of Science and Technology, India and by the project M~1508 of the Austrian Science Fund (FWF). 
DG and MR were supported by the START project Y~435-N16 of the FWF and the FWF projects I~952-N16, I~1030-N27 and P~27182-N27.

\bibliographystyle{fullsort}
\bibliography{review}

\end{document}